\documentclass[aps,prl,twocolumn,superscriptaddress,showpacs]{revtex4}
\usepackage{graphicx}

\usepackage{dcolumn}

\usepackage{bm}
\usepackage{color}

\begin{document}
\title{Terahertz plasma wave generation in ultra-short-channel Field Effect Transistors: theory vs experiment}

\author{M. V. Cheremisin, G. G. Samsonidze}

\affiliation{A.F.Ioffe Physical-Technical Institute,
St.Petersburg, Russia}

\begin{abstract}
Taking into account both the scattering and the velocity
saturation of carriers, we examine the "shallow-water" instability
of the two-dimensional electron gas in a field effect transistor.
It is shown that both the scattering (which is analogous to
friction in a shallow-water channel) and the carrier velocity
saturation lead to damping of the plasma wave instability.
Threshold diagram of instability is calculated. The actual device
parameters required for observation of plasma wave generation are
compared with those reported in recent sub-terahertz emission
experiments.
\end{abstract}


\maketitle

\section{\label{sec:Introduction}Introduction}
Recently, the interest in the plasma wave generation mechanism [1]
in an ultrashort-channel Field Effect Transistor (FET) has
resumed. As it was demonstrated in Ref.[1], the behavior of
high-density two-dimensional electron gas (2DEG) can be described
by equations analogous to the hydrodynamic equations for shallow
water, the plasma waves being similar to waves in this medium.
With the carrier scattering neglected, it is shown that asymmetric
boundary conditions with a fixed voltage at the source and fixed
current at the drain lead to an instability of steady state with a
dc current. The FET channel can be regarded as a resonator, with a
plasma-wave generation mechanism similar to that known for
self-excitation of jets and organ pipes. Both the eigenfrequencies
and the instability increment have been found for zero-dissipation
electrons. In Refs.[2,3] a strongly nonlinear instability mode, in
which shock waves are formed in the channel, was analyzed. In
Ref.[4], the instability increment and threshold were calculated
for the realistic case of nonzero carrier scattering at arbitrary
currents and electron momentum relaxation times. The scattering
results in the narrowing of the instability region, with the
increment decreasing. The instability vanishes at a certain
critical magnitude of friction. Computer simulations [4] provide a
strong support for theoretical predictions. The plasma wave
instability in short-channel FETs is very important for
application of these devices as high-power sources and detectors
in the terahertz (THz) frequency range[5]-[10](see also [11]). The
sub-THz emission recently observed in short-channel FETs [12,13]
was attributed to the plasma wave generation mechanism [1].

In the present paper, we analyze the plasma wave instability in
the presence of both the scattering and the velocity saturation of
carriers. We plot the instability diagram for device parameters
required for plasma wave generation. We demonstrate that the 2DEG
samples studied in [12,13] have an insufficient carrier mobility
and, hence, the observed sub-THz emission cannot be understood in
terms of the plasma wave generation scenario.

\section{\label{sec:Analytical approach}Analytical approach}
\subsection{\label{sec:Steady state}Steady state}

According to Ref.[1], the high-density 2DEG in an ultra-short
channel FET can be described by the following equations analogous
to the hydrodynamic equations for shallow water:

\begin{eqnarray}
\frac{\partial U}{\partial t}+\frac{\partial (VU)}{\partial x}=0,
\label{continuity} \\
\frac{\partial V}{\partial t}+\frac{\partial }{\partial x}\left(
\frac{V^{2}}{2}+\frac{eU}{m}\right)+\frac{V}{\tau}=0.
\label{Euler}
\end{eqnarray}
Here, the voltage swing, $U=U_{gc}-U_{T}$, corresponds to a
shallow water level, $U_{gc}$ is the local gate-to-channel
voltage, $U_{T}$ the threshold voltage, $V$ the local electron
flux velocity, $m$ the effective mass, and $\tau$ the momentum
relaxation time due to collisions of electrons with phonons and
(or) impurities. Eq.(1) is the continuity equation, in which the
Shockley relation for the gradual channel approximation, $N=CU/e$,
is taken into account. Here, $N$ is the surface electron
concentration, and $C$ is the capacitance per unit area. Note that
there is no direct hydrodynamic analogy for the friction term
$\frac{V}{\tau}$ in the Euler equation specified by Eq.(2).

It has been shown [1] that in the absence of friction($\tau
\rightarrow \infty$), which is further named the "clean limit,"
the steady electron flow with a constant $U,V$ is unstable against
small perturbations $\delta V, \delta U \sim \exp(-i\omega t)$
under the following boundary conditions:
\begin{equation}
U(0,t)=U_{s},\qquad U(l,t)V(l,t)=j/WC, \label{boundary conditions}
\end{equation}
where $U_{s}$ is the fixed source($x=0$) potential, $j$ the
current fixed at the drain ($x=l$), and $W$ the gate width. In the
steady state, $U=U_{s},V=V_{s}=j/(WCU_{s})$. The real and
imaginary parts of $\omega$( see Figs.1, 2) are given by [1]
\begin{equation}
\omega'_{n}=\frac{S}{2l}|1-M^{2}|\pi n,
\omega''=\frac{S}{2l}(1-M^{2})\ln \left | \frac{1+M}{1-M}\right |,
\label{Dyakonov}
\end{equation}
where  $M= V/S$ plays the role of the Mach number in
hydrodynamics, $S=\sqrt{eU/m}$ is the local plasma wave velocity,
and $n$ is an odd integer for $|M|<1$ and even integer for
$|M|>1$. Actually, the instability is related to the plasma wave
amplitude enhancement caused by certain boundary reflections[1].
Two different plasma waves with wave vectors $k_{1,2}=\pm
\omega/(S(1 \pm M))$ propagate down- and upstream, respectively.
The argument of the logarithm in Eq.(4) contains the round-trip
gain factor, $(1+M)/(1-M)$, which is the product of the reflection
coefficients at two boundaries. The plasma wave propagates
downstream during the time, $l/(S(1+M))$, whereas the back
propagation takes a longer time, $l/(S(1-M))$. When $M \rightarrow
1$, the round-trip time of the plasma wave becomes infinite, and,
therefore, both the proper frequency and the instability increment
vanish ( see Figs.1,2 ).

The friction of electron flow results in a spatial dependence [14]
of both the voltage and the velocity. In this case, the
instability problem for small perturbations superimposed on the
steady state flow $V,U$ becomes extremely difficult. Fortunately,
this problem can be solved by the high-order mode method suggested
in Ref.[4]. This approach makes it possible to deduce both the
instability increment and the threshold for arbitrary currents and
friction strengths. As expected, the scattering leads to
instability damping. At a certain critical friction, the
instability vanishes.

The special interest of the present paper is in the friction term
which has always been assumed to contain a constant momentum
relaxation time $\tau$ previously [14,4]. We argue that the drift
velocity in real 2D systems at high electric fields approaches a
certain saturation value $V_{sat}$. In the conventional FETs just
the drift velocity saturation (DVS) mechanism determines the
saturation current. In order to account for the possible influence
of the DVS effect on the instability, we further investigate the
realistic model with the momentum relaxation time in Eq.(2)
replaced as
\begin{equation}
\tau \rightarrow \frac{\tau}{1+\mu E/V_{sat}}, \label{tau}
\end{equation}
where $\mu$ is the low-field mobility of the 2D system, and
$E=-\frac{dU}{dx}$ is the electric field. We further demonstrate
that the DVS effect leads to an extra damping of the plasma wave
instability.

Using Eqs.(1,2), we obtain for the steady state
\begin{eqnarray}
j=WCVU=const, \nonumber \\
\frac{\partial M}{\partial \eta}=\frac{3}{2}\frac{\gamma }{
M_{s}^{1/3}}\frac{M^{7/3}}{\phi(M,p)}, \label{steady state}
\end{eqnarray}
where $\eta=x/l$ is the dimensionless coordinate,
$\gamma=l/(S_{s}\tau)$ the friction parameter, and $S_{s}$ the
plasma wave velocity at the source. Then, $\phi(M,p)=1-M^{2}-psM$,
where $s=S/S_{s}=(M_{s}/M)^{1/3}$  is the dimensionless local
plasma wave velocity, and $M_{s}$ is the Mach number at the
source. Then, we introduce the dimensionless parameter
$p=\frac{S_{s}}{V_{sat}}$ associated with the DVS effect. Previous
studies [14,4] correspond to the $p=0$ case.

We are further interested only in the case of a subsonic flow
case, when $0<M \leq 1$. According to Eq.(6), at finite $\gamma$
the Mach number increases from the initial value $M_{s}$ at the
source to some drain value $M_{d}=M(1)$. Note that at fixed $p$
the increasing solution to Eq.(6) is possible when
$\phi(M_{s},p)>0$ or $M_{s}<M_{s}^{max}(p)=\sqrt{p^{2}/4+1}-p/2$.
If $M_{s}<M_{s}^{max}$, the electron flow velocity saturates at
the drain, where the electric field is the strongest (i.e., $E
\sim \frac{dM}{d\eta} \rightarrow \infty $ ), as the expression in
the denominator in the right-hand part of Eq.(6) goes to zero. The
Mach number at the drain cannot exceed a certain value
$M_{d}^{sat} \leq 1$ specified by the condition $\phi(M_{d},p)=0$.

It is instructive to compare our results with those provided by he
hydrodynamic "choking" saturation [14] and the conventional DVS
mechanism separately. Without DVS (i.e. $p=0$), we reproduce the
"choking" criterion [14] as $M_{d}^{sat}=1$. In the conventional
hydrodynamics, the above condition corresponds to a subsonic gas
flow which cannot evaluate to a supersonic one. In the opposite
DVS case, (see, for example, Ref.[15]) the current is saturated
when the condition $1-psM=0$ is satisfied at the drain contact. We
emphasize that both the mechanisms combined (as in Eq.(6)) result
in a faster current saturation, compared with the case when each
of them is considered separately.

The solution to Eq.(6) determines the steady-state spatial
dependence of the Mach number $M(\eta)$:
\begin{equation}
f(M_{s})-f(M)=\frac{2\gamma \eta}{M_{s}^{1/3}},
\label{M_distribution}
\end{equation}
where $f(M)=M^{-4/3} +2M^{2/3}-2pM_{s}^{1/3}M^{-2/3}$. At a fixed
friction strength $\gamma$, the Mach numbers $M_{s}$ and $M_{d}$
are related by
\begin{equation}
f(M_{s})-f(M_{d})=\frac{2\gamma}{M_{s}^{1/3}}.
\label{Md vs Ms}
\end{equation}
Substituting $M_{d}=M_{d}^{sat}$ into Eq.(8), we obtain the
saturation threshold $\gamma_{sat}(M_{s},p)$, represented in Fig.3
by dotted lines, for different values of the DVS parameter $p$. As
expected, the saturation current($ \sim M_{s}$) at fixed $\gamma$
decreases as the DVS parameter $p$ increases.

\subsection{\label{sec:Instability}Instability}

We now study the stability of the steady-state flow by inspecting
the temporal behavior of  small perturbations $V_{1}(\eta),U_{1}
(\eta)$ superimposed on a steady flow with a velocity $V(\eta)$
and channel potential $U(\eta)$. Thus, we take
\begin{eqnarray}
V(\eta,t)=V(\eta)+V_{1}(\eta)exp(-i\omega t), \nonumber \\
U(\eta,t)=U(\eta)+U_{1}(\eta)exp(-i\omega t). \label{perturbation}
\end{eqnarray}
Equations (1),(2),(3), linearized with respect to $V_{1}$ and
$U_{1}$, yield:
\begin{eqnarray}
-i\Omega \gamma v_{1} + \frac{d}{d \eta}(v_{0}v_{1}+u_{1})-p(v_{1}u'_{0}+v_{0}u'_{1})+\gamma v_{1}=0 , \nonumber \\
-i\Omega \gamma u_{1} + \frac{d}{d \eta}(u_{0}v_{1}+u_{1}v_{0})=0. \nonumber \\
u_{0}(0)=0, u_{0}(1)v_{1}(1)+u_{1}(1)v_{0}(1)=0.
\label{linearizing}
\end{eqnarray}
where $v_{1}= V_{1}/S_{s}, v_{0}= V(\eta)/S_{s}, u_{1}=
U_{1}/U_{s}, u_{0}= U(\eta)/U_{s}$. Then,
$\omega=\Omega/\tau=\Omega \gamma S_{s}/l$, where $\Omega$ is the
dimensionless frequency. The complex frequency $\Omega=\Omega' +
i\Omega''$ can be determined by solving these equations. A
positive imaginary part $\Omega'' >0$ corresponds to instability.

\begin{figure}[t]
%h=here, t=top, b=bottom, p=separate figure page
\begin{center}
 \leavevmode
\includegraphics[scale = 0.25,angle=0,keepaspectratio=true,width=2.75in]{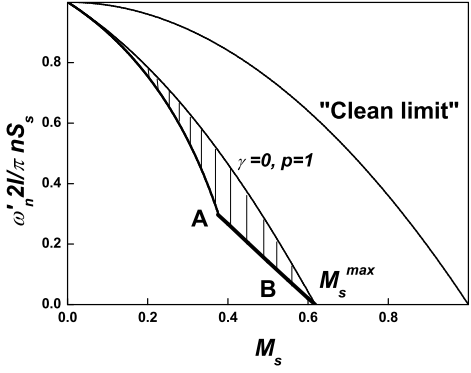}
 %\epsfxsize=3in
 %\epsfbox {tiltfigures2V2eps.eps}
\end{center}
\caption{Dimensionless frequency $\frac{\omega'_{n}2l}{\pi n
S_{s}}$ vs the Mach number at the source for the "clean limit"[1]
at $p;\gamma=0$. The shadowed area corresponds to that in the
instability threshold diagram (see Fig.3) and shows all the
frequencies that can be generated ( $\omega''>0$ ) at fixed $p=1$
and $0<\gamma<0.32$. The bold line A-B shows the frequency
$\omega'(M_{s})$ in the saturation mode (see the corresponding
line in Fig.3). The point B corresponds to $M_{s}^{max}(1)=0.62$.}
\label{Fig1}
\end{figure}

\begin{figure}[t]
%h=here, t=top, b=bottom, p=separate figure page
\begin{center}
 \leavevmode
\includegraphics[scale = 0.25,angle=0,keepaspectratio=true,width=2.75in]{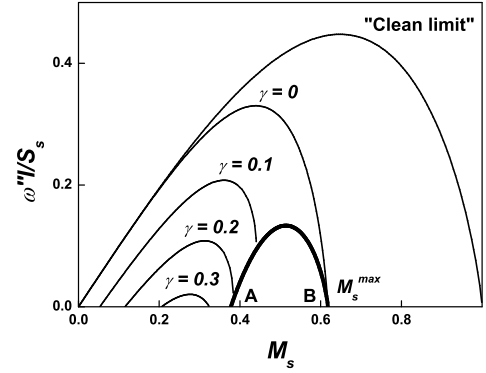}
 %\epsfxsize=3in
 %\epsfbox {tiltfigures2V2eps.eps}
\end{center}
\caption{Dimensionless increment of instability $\omega''l/S_{s}$
vs Mach number at the source for the: "clean limit"[1] at
$p;\gamma=0$, 2DEG with DVS($p=1$) and different values of the
friction parameter $\gamma=0;0.1;0.2;0.3$. The bold line A-B shows
the increment of instability in the saturation mode (represented
by A-B line in Fig.3). The point B corresponds to
$M_{s}^{max}(1)=0.62$.} \label{Fig2}
\end{figure}

It is instructive to introduce the current perturbation
$w=u_{0}v_{1} +u_{1}v_{0}$. With the help of a simple relationship
$v_{0}=(M_{s}M^{2})^{1/3}$ specified by Eq.(6), Eq.(10) can be
re-written as follows:
\begin{eqnarray}
w''\phi(M)+w'\frac{\gamma M}{s}\left (
2i\Omega-\frac{1+2M^{2}}{\phi(M)}\right )+ \\ w\frac{\Omega
\gamma^{2}}{s^{2}} \left( \Omega+i\frac{1+M^{2}}{\phi(M)} \right )=0, \nonumber \\
w'(0)=0, w(1)=0, \label{current perturbation}
\end{eqnarray}

We remind that the scale of the steady state nonuniformity is the
channel length. The wavelengths of the lower-order modes of the
plasma oscillations are on the same order of magnitude. In
contrast, the higher-order modes have a considerably shorter
scale. As stated in Ref.[4], the higher modes are more sensitive
to the steady state nonuniformity than the fundamental and
low-order modes. Since the increment of instability is strongly
affected by the boundary-reflection coefficients, it would be
expected that the higher-order modes are amplified more
effectively. Based on such a qualitative argumentation, we analyze
Eq.(11) within the higher-order mode approximation.

\begin{figure}[t]
%h=here, t=top, b=bottom, p=separate figure page
\begin{center}
 \leavevmode
\includegraphics[scale = 0.25,angle=0,keepaspectratio=true,width=2.75in]{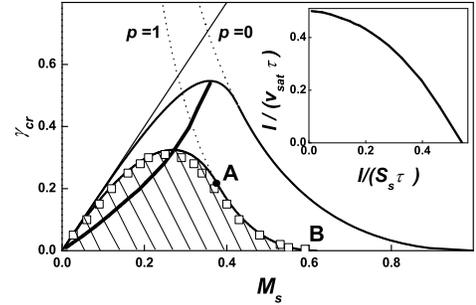}
 %\epsfxsize=3in
 %\epsfbox {tiltfigures2V2eps.eps}
\end{center}
\caption{Diagram of instability threshold. The critical value of
the friction parameter, $\gamma_{cr}$ vs Mach number at source
(bold lines) and saturation threshold $\gamma_{sat}$ (dotted
lines) for $p=0$ (see Ref.[4]) and $p=1$. The thin line is the
asymptote $\gamma_{cr} =2M_{s}$. The increment of instability
along the line A-B is shown in Fig.2. The point B corresponds to
$M_{s}^{max}(1)=0.62$. The ultra-bold line represents the maximum
of the instability threshold $\gamma_{max}({M_{s}})$ for $0<p<
\infty $. The data obtained by numerical simulations based on
Eqs.(1,2) are shown by squares. Inset: FET parameters $\gamma$ vs
$\gamma*$ required for plasma wave generation.} \label{Fig3}
\end{figure}

The spatial dependence of the Mach number $M(\eta)$ is a
monotonous function of the coordinate. Consequently, we can
represent the current perturbation $w$ as a function of the Mach
number $w=w(M)$. Moreover, following Ref.[4] we search for the
solution in the form $w=F \exp(g)$, where
$g=-\frac{2}{3}(i\Omega+5/4)\ln(M)$. Finally, we use Eq.(6) and,
then, change the independent variable from $\eta$ to $M$, to
rewrite Eq.(11) as follows:

\begin{eqnarray}
\frac{9}{4}M^{4}F''+ \left [ \Omega^{2}(1-p s M)+i\Omega
+\frac{5}{16}M^{2} \right ]F=0. \nonumber \\
w'(M_{s})=0, w(M_{d})=0. \label{Basic F-equation}
\end{eqnarray}
where we now denote by $F', F'', w', w''$  the derivatives with
respect to $M$. It is noteworthy that Eq.(13) is exact. Solutions
to this equation with given boundary conditions exist only for
certain complex frequencies $\Omega$ (eigenvalues of the problem).
Note the analogy between Eq.(13) and one-dimensional Schrödinger
equation, in which the complex frequency $\Omega$ plays the role
of energy in the quantum mechanics. In general, Eq.(13) can be
solved only numerically. However, an analytical solution can be
found using the high-order mode method [4] which corresponds to
the quasi-classical approximation in quantum mechanics. This
approach is valid when $|\Omega| \gg 1$, and, therefore, the
imaginary part of $\Omega$ is small as compared with its real
part, $\Omega' \gg \Omega''$. We now search for a solution to
Eq.(13) in the form $F=\frac{A}{\sqrt{k}}\exp(-i\int k d M)$,
where $A$ is a constant, and $k(M)$ is the dimensionless wave
vector. Finally, we obtain the dispersion equation as
\begin{equation}
k_{1,2}=\pm \frac{2}{3M^{2}} \left( \Omega
\sqrt{1-psM}+\frac{i}{2\sqrt{1-psM}} \right ).
\label{k-vectors}
\end{equation}
Here we keep the main corrections with respect to powers of
$\Omega$, and, therefore, neglect the term $\frac{5}{16}M^{2}$ in
Eq.(13). The wave vectors $k_{1}, k_{2}$ correspond to the plasma
waves propagating down- and upstream, respectively. Searching the
solution to Eq.(13) in the form $F=\sum \limits_{l=1,2}
\frac{A_{l}}{\sqrt{|k_{l}|}}\exp(-i\int k_{l} d M)$, with the
given boundary conditions, we can easily find:
\begin{equation}
\exp \left( i \int \limits_{M_{s}}^{M_{d}}(k_{2}-k_{1})dM
\right)=R_{s}^{I}R_{d}^{I}
\label{instability condition}
\end{equation}
where $R_{s}^{I}=\frac{\sqrt{1-psM}+M_{s}}{\sqrt{1-psM}-M_{s}}$,
$R_{d}^{I}=-1$ are the current amplitude reflection coefficients
from the source and drain boundaries, respectively, and
$R_{s}^{I}R_{d}^{I}$ is the round-trip gain factor. Eq.(15) allows
finding of both the real and imaginary parts of $\Omega$ as
follows:
\begin{eqnarray}
\Omega'_{n}=\frac{3\pi n}{4} \frac{1}{ \frac{(1-psM)^{3/2}}{M} |_{M_{s}}^{M_{d}}}, \label{frequency} \\
\Omega''= \frac{ -\frac{\sqrt{1-psM}}{2M}(1+2psM)
|_{M_{s}}^{M_{d}} + \frac{3}{4}\ln (R_{s}^{I}R_{d}^{I}) } {
\frac{(1-psM)^{3/2}}{M} |_{M_{s}}^{M_{d}}}. \label{increment}
\end{eqnarray}

We emphasize that, in the "clean limit" [1], i.e., at
$\gamma;p=0$, Eqs.(16,17) coincide with those specified by Eq.(4).
Using of Eq.(17) and the relationship between the Mach numbers at
the source and drain given by Eq.(8), we can easily obtain the
dependence of the instability increment $\omega''$ on Mach number
at the source $M_{s}$ for arbitrary friction $\gamma$ and DVS
strength $p$. Both the friction and the DVS effect result in
instability damping (see Fig.2). Firstly, DVS leads to narrowing
of the instability range. For example, at $p=1$ and $\gamma=0$ the
instability exists within the range $0<M_{s}<M_{s}^{max}(1)=0.62$,
which is narrower than that $0<M_{s}<1$ in the "clean limit".
Secondly, at certain DVS strength the enhancement of friction
suppresses the instability (see the curves for $p=1$ in Fig.2).
The region of instability narrows as well. Actually, in the case
of $p=1$ (Fig.2), the instability exists when $\gamma<0.32$. This
condition is more stringent as compared with the analogous
$\gamma<0.54$ in the absence ($p=0$) of the DVS mechanism.

Substituting $\Omega''=0$ into Eq.(17) and using Eq.(8), we plot
in Fig.3 the instability threshold $\gamma_{cr}(M_{s},p)$ for two
different DVS strengths $p=0;1$. In addition, the same figure
shows by dotted lines the corresponding dependencies for the
saturation threshold $\gamma_{sat}(M_{s},p)$. At a certain point
(denoted by "A" for the $p=1$ case) both the threshold curves
merge, and, thus give the threshold diagram of instability, marked
by the shadowed area in Fig.3. The threshold diagram is confined
between two asymptotes. One of these (thin line in Fig.3), given
by $\gamma_{cr} =2M_{s}$, is the same for both $p=0$ and $p=1$
cases. Actually, this is the result obtained previously in Ref.[1]
for the case of a small friction, at which the spatial
distribution of $M$ is only slightly nonuniform. Then, the
threshold diagram is bounded at high $M_{s}$ by the saturation
threshold curve $\gamma_{sat}(M_{s},p)$. Note that, within a part
of the latter asymptote (for the $p=1$ case, the bold line A-B in
Fig.3), the increment of instability is positive, as shown in
Fig.2. In what follows, we explain the above discrepancy by the
uncertainty of our high-order mode method.

Arguing that Fig.3 represents the instability threshold, we, in
addition, determine $\gamma_{cr}$ by numerical simulations based
on Eqs.(1,2) and Eq.(3), using the previously described numerical
method [2]. The results obtained (shown by squares in Fig.3)
coincide with our theoretical curve within the error of the
numerical method. As the numerical simulations give the threshold
for the most unstable modes, this coincidence with the
analytically obtained results, clearly demonstrates the validity
of our approach.

We now present our central result, concerning the device
parameters required for a possible observation of the instability.
First, for certain $p$ we find the maximum of the instability
threshold $\gamma_{max}(p)$ (the corresponding dependence
$\gamma_{max}(M_{s})$ is represented by the ultra-bold line in
Fig.3). Then, the general instability condition reads
$\gamma<\gamma_{max}(p)$. It will be recalled that
$p=\frac{\gamma*}{\gamma}$, where we introduce the dimensionless
parameter $\gamma*=\frac{l}{v_{sat}\tau}$ related to the DVS
effect. With the above notation, the instability criterion yields
\begin{equation}
\gamma* > \gamma_{max}^{-1}(\gamma) \gamma, \label{FET_parameters}
\end{equation}
where $\gamma_{max}^{-1}(x)$ is the inverse function. In the inset
in Fig.3, we plot the diagram specified by Eq.(18). The values of
the FET parameters, confined by the bold curve, correspond to an
instability. Another condition is related to the applicability of
the hydrodynamic model suggested in Ref.[1]. In actual fact, 2D
electrons can be considered a fluid when the mean free path
$\lambda \sim \sqrt{N}$, associated with electron-electron
collisions, is shorter than the channel length. This requirement
can be easily transformed to $\gamma <
\frac{l^{2}}{\mu}\sqrt{C/m}$.

The final question remaining to be resolved is whether the finite
scattering and the DVS effect can influence the proper and the
higher-order-mode frequencies of the instability. With the help of
Eq.(16), we calculate the nth-order-mode frequency
$\omega'_{n}(M_{s},p,\gamma)$. At a fixed $p=1$ in Fig.1, we plot
the shadowed area (which corresponds to that in Fig.3) which shows
all the possible plasma frequencies when the instability increment
is positive $\Omega''>0$. At a fixed current $j \sim M_{s}$ and a
finite $p$, the proper frequency and its harmonics may be much
lower than those expected in "clean limit" (see Eq.(4)). The
difference between two cases becomes more pronounced in the
saturation mode, at $M_{s} \leq M_{s}^{max}$. We conclude that the
DVS effect results in a strong depression of the plasma wave
generation frequency.

We are now going to justify the applicability of the
high-order-mode approach[4], developed in the present paper. We
first compare the data presented in Figs.1,2, and, then conclude
that our basic assumption $\Omega'>\Omega''$ is, surprisingly,
valid for the proper plasma wave mode $n=1$ as well. This is,
however, not the case in the saturation mode, at $\Omega
\rightarrow 0$, which seems to be the reason for a minor deviation
between the theory and the results of our numerical simulations
(see threshold diagram in Fig.3).
\begin{figure}[t]
%h=here, t=top, b=bottom, p=separate figure page
\begin{center}
 \leavevmode
\includegraphics[scale = 0.25,angle=0,keepaspectratio=true,width=2.75in]{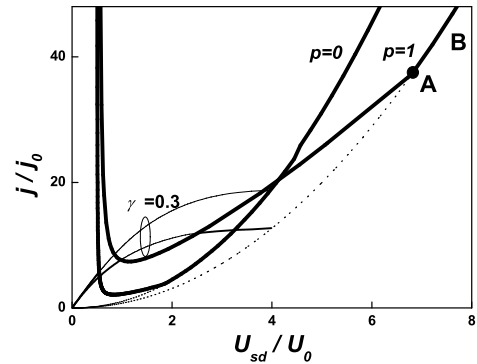}
 %\epsfxsize=3in
 %\epsfbox {tiltfigures2V2eps.eps}
\end{center}
\caption{Instability region in the I-V characteristic. The I-V
curves are plotted for $\gamma=0.3$, where the upper(lower) curve
corresponds to $p=0$ and $p=1$, respectively. For $p=0$ and $p=1$
cases the bold (dotted) line (corresponding to those in Fig.3)
indicates the instability (saturation) threshold respectively. The
unstable parts of the current-voltage characteristics are
indicated by thin lines. The line A-B corresponds to that shown in
Fig.3} \label{Fig4}
\end{figure}

Let us represent the instability threshold shown in Fig.3 in
current-voltage( I-V ) plane of the transistor. In terms of the
Mach number formalism, the current and the source-drain voltage
swing yield
\begin{equation}
\frac{j}{j_{0}}=\gamma^{-3} M_{s},
\frac{U_{sd}}{U_{0}}=\gamma^{-2}(1-(M_{s}/M_{d})^{2/3}),
\label{IV}
\end{equation}
where we use the units $j_{0}=WCm(l/\tau)^{3}/e$ and
$U_{0}=m(l/\tau)^{2}/e$, introduced in Ref. [14]. For a fixed
$M_{s},\gamma, p$, we can calculate $M_{d}$ specified by Eq.(8)
and then use Eq.(19) to find the steady-state I-V characteristics
$j(U_{sd})$. In Fig.4 we plot I-V characteristic for certain
values $p=0;1$. In addition, the instability threshold curves
$\gamma_{cr}(M_{s},p)$ can be transformed to those
$j_{cr}(U_{sd})$ in the I-V plane. Furthermore, the dotted curves
(corresponding to $\gamma_{sat}(M_{s},p)$ in Fig.3) relate the
saturation current to saturation source-drain voltage. For the
$p=1$ case, both the threshold curves, which merge at point A,
confine the instability region in the I-V plane. The parts of the
current-voltage characteristics lying within this region
correspond to unstable states (thin lines in Fig.4). For
$\gamma>0.32(p=1)$ and $\gamma>0.54(p=0)$, the current-voltage
characteristics are always stable. Note that the saturation
current at finite $p$ is lower than that that in absence of the
DVS effect.

\subsection{\label{sec:Theory vs experiment}Theory vs experiment}

We now provide a detailed analysis of the data obtained in a
THz-emission experiment[12] for an ultrashort-channel InGaAs-based
FET. In this device, the source-drain separation of $1.3\mu$m was
made much longer than the gate length $l=60$nm. Thus, the total
series resistance $R$ associated with the gate-source and
gate-drain access regions may affect the I-V characteristics of
the device. Indeed, the voltage bias between the source and drain
ends of the gated part of the channel $U_{sd}$ is related to the
external (hereafter marked by prime) source-to-drain terminal
voltage by $U'_{sd}=U_{sd}+jR$. Then, for the actual case of a
symmetric FET (see inset in Fig.5) the swing voltage $U_{s}$ is
related to the external gate-to-source voltage by
$U_{s}=U'_{gs}-U_{T}-jR/2$. We further use the simplest channel
pinch-off model, for which the I-V dependence of the device is
given by $j=\beta U_{sd}(U_{s}-U_{sd}/2)$, where $\beta=WC\mu/l$
is the I-V steepness. Assuming that the actually measured I-V
characteristic is expressed in terms of terminal voltages
$U'_{gs}$, $U'_{sd}$, we derive I-V dependence as follows
\begin{equation}
j=\beta\frac{U'_{sd}(U'_{s}-U'_{sd}/2)}{1+\beta
R(U'_{s}-U'_{sd}/2)}.
\label{pinch-off+}
\end{equation}
where $U'_{s}=U'_{gs}-U_{T}$. We now analyze the I-V transfer
characteristics measured[12] at a fixed source-drain bias
$U'_{sd}=5$mV. According to Eq.(20), if the applied gate-to-source
terminal voltage $U'_{gs}$ exceeds a certain threshold value
$U_{T}+U'_{sd}/2$, the current is nonzero. Then, at high
$U'_{gs}$, the current saturates, being limited only by the only
series resistance, and, hence $j=U'_{sd}/R$. In Fig.5, we
reproduce the observed FET transfer characteristic (see inset in
Fig.1 of [12]) and then put on the same plot our best-fit data.
Note that Eq.(20) is, in fact, valid only in the gradual Shockley
approximation, and, therefore, the subthreshold mode
$U'_{gs}-U_{T}<0$ is not analyzed in Fig.5. The device parameters
obtained are as follows: $\beta=0.66 \frac{1}{\Omega V},
R=14\Omega$, and $U_{T}=-147$mV. For $\epsilon=12.7$,
gate-to-channel separation $d=17$nm, and gate width $W=50\mu$m,
the FET mobility is $\mu=1200$cm$^{2}$/Vs. The corresponding
momentum relaxation time is $\tau=2.9 \times 10^{-14}$s.

\begin{figure}[t]
%h=here, t=top, b=bottom, p=separate figure page
\begin{center}
 \leavevmode
\includegraphics[scale = 0.25,angle=0,keepaspectratio=true,width=2.75in]{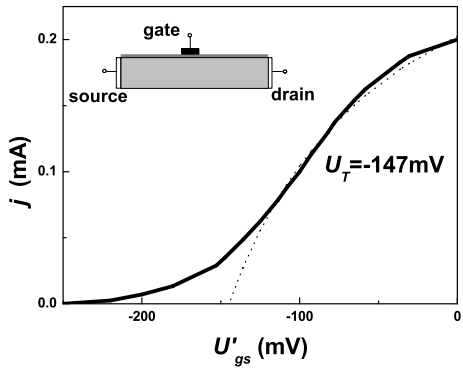}
 %\epsfxsize=3in
 %\epsfbox {tiltfigures2V2eps.eps}
\end{center}
\caption{I-V transfer characteristic ( bold line ) at
$U'_{sd}=5$mV, reproduced from Ref.[12]. Dotted line represents
the fit with Eq.(20). Inset: schematic of the short-channel FET.}
\label{Fig5}
\end{figure}

The next step of our analysis is concerned the FET I-V
characteristic [12] at a fixed gate-to-source voltage (in the
experiment $U'_{gs}=0$, and, therefore, $U'_{s}=-U_{T}$). At a
fixed $U'_{s}$ the current is linear in the source-to-drain bias
$j=\frac{U'_{sd}}{R_{ch}+R}$, where $R_{ch}=(\beta U'_{s})^{-1}$
denotes the channel resistance. At a certain source-to-drain bias,
the device current saturates (channel pinch-off) when
$dj/dU'_{sd}=0$. Both the saturation current and voltage are given
by
\begin{eqnarray}
j_{sat}=2 \beta (U'_{s})^{2} \left(\frac{\sqrt{1+r}-1}{r} \right )^{2}, \nonumber \\
U'_{sd}|^{sat}=2U'_{s}\frac{1+r-\sqrt{1+r}}{r},
\label{saturation+}
\end{eqnarray}
where $r=R/R_{ch}$. Without the series resistance $R$, Eq.(21)
reproduces the conventional channel pinch-off saturation at
$j_{sat}=\beta U_{s}^{2}/2$, $U_{sd}^{sat}=U_{s}$. By contrast, at
a nonzero $R$, the saturation current decreases and the saturation
voltage increases as compared with the $R=0$ case. For the I-V
characteristic reported in [12], we have $U'_{s}=147$mV, and,
therefore, the channel resistance is $R_{ch}=11 \Omega$. The total
resistance $R+R_{ch}=14+11=25$ is comparable with the value of
$29\Omega$, obtained in [12] from the linear part of the I-V
curve. Then, we estimate both the saturation current
$j_{sat}=4.4$mA and voltage $U'_{sd}|^{sat}=180$mV, which are,
however, somewhat low than those obtained in the experiment.

We now discuss the most intriguing result[12] concerning the
irradiation $f \sim 0.4$THz, observed at certain current $j=4.5$mA
and voltage $U'_{sd}=200$mV. The above emission threshold
corresponds to $U_{sd}=U'_{sd}-jR=70$mV and effective voltage
swing $U_{s}=U'_{gs}-U_{T}-jR/2=116$mV. In contrast to Ref.[12],
we conclude that the observed emission threshold precedes the
saturation mode of the transistor because $U_{sd}<U_{s}$. Our
conclusion can be further confirmed by analyzing the actual shape
of the I-V characteristic reported in Ref.[12]. In general, the
output drain conductance vanishes in the saturation mode, i.e.
$\frac{dj}{dU'_{sd}}=0$, irrespective of the actual mechanism
(channel pinch-off, DVS mechanism based on smooth $\tau(E)$
dependence, "chocking"). Obviously, this is not the case for the
THz-emission threshold point reported in Ref.[12].

\begin{table}
\caption{\label{tab:table1}Experiment vs theory}
\begin{ruledtabular}
\begin{tabular}{cccccccc}
2DEG       &$l|d$, &$R$, &$U_{T}$,  &$\mu$,    &$S_{s}|V_{sat}$   &$\gamma|\gamma*$ \\
                           & nm   &$\Omega$&V   &$\frac{\text{cm}^{2}}{\text{Vs}}$&$\times 10^{5}\frac{m}{s}$&&          \\
 \hline      InGaAs[12]&$60|17$  &14&-0,147&1200      &$7.0|3.2$           &$\bf{3.0}|\bf{6.6}$      \\
              $m*=0.042$ &&&&& \\
          GaN[12] &$1500|25$&62&-2.8&1700     &$11.3|3.5$            &$\bf{6.3}|\bf{20}$        \\
           $m*=0.22$& & &&&\\

\end{tabular}
\end{ruledtabular}
\end{table}

We now verify whether the device parameters $\gamma,\gamma*$ match
the limits( see Fig.4, inset ) required for plasma wave
instability to occur. For $m=0.042m_{0}$, we find the plasma wave
velocity $S_{s}=\sqrt{eU_{s}/m}=7.0 \times 10^{5}$m/s and the 2D
density $N_{s}=\frac{\epsilon_{0}\epsilon U_{s}}{d e}=3.5 \times
10^{11}$cm$^{-2}$. We estimate the friction strength as
$\gamma=\frac{l}{S_{s}\tau}=3.0$. Then, with the textbook value
for the saturation velocity $V_{sat}=3.2 \times 10^{5}$m/s, we
obtain $p=2.2$, hence $\gamma*=6.6$. Both the parameters $\gamma,
\gamma*$ exceed the limits ( see Fig.4, inset ) required for
instability. Hence, the observed sub-THz emission [12] cannot be
attributed to the plasma wave instability scenario. Moreover, our
step-by-step analysis (see Table 1) of the data for a
$1.5\mu$m-channel GaN FET[12] shows that the GHz-emission observed
does not originate from the "shallow-water" mechanism, either.

\subsection{\label{sec:Conclusion}Conclusion}

We have shown that both the scattering and the DVS effect depress
the frequency and the instability increment. The experimental
results reported in Refs.[12],[12] for sub-THz emission in
ultra-short channel FETs cannot be explained in terms of the
plasma instability scenario[1].

\begin{acknowledgments}
This study was supported by Large Scale Facility program (
Weizmann Institute, Israel, grant No.HPRI-CT-2001-00114 ) and
Russian Foundation for Basic Research ( grant No.03-02-17588).
\end{acknowledgments}

\end{document}